# Modeling the geographical studies with GeoGebra-software

**Ionica Soare**
"Dunărea de Jos" University of Galaţi, România
**Carmen Antohe**
National College „N. Bălcescu" of Brăila, România

**ABSTRACT.** The problem of mathematical modeling in geography is one of the most important strategies in order to establish the evolution and the prevision of geographical phenomena. Models must have a simplified structure, to reflect essential components and must be selective, structured, and suggestive and approximate the reality. Models could be static or dynamic, developed in a theoretical, symbolic, conceptual or mental way, mathematically modeled. The present paper is focused on the virtual model which uses GeoGebra software, free and available at www.geogebra.org, in order to establish new methods of geographical analysis in a dynamic, didactic way.

## 1. Materials and Methods

Due to the transition to new pedagogical technologies, the practice demands the introduction of the dynamic strategies into the pedagogical activity. In this context a training-based on the use of a personal computer and software giving the possibility to use the IT and AEL methods seems attractive. On this reason one's need to use specialized software or free platform seems to be essential. One of this software, that will be proposed to be used, is GeoGebra software. Some knowledge of math and Euclidian geometry will be necessary. The basic advantages of using GeoGebra in a school, in an interdisciplinary geography course are the following: individualization in teaching and training; providing independent activity for pupils; possibility of analysis helping to strengthen and to master the material; possibility of working with an algorithmic method, which produces active development of logic thinking [Ant09]. At the given stage one must conduct work both in strategy and in methodological directions. The cooperation between experts (math teacher, geography teacher) creates and develops corresponding techniques of geometry and geographical representation, in order to realize electronic materials for educational purpose and make them more expedient pedagogically.

## 2. Modeling a geometric topology with GeoGebra

The statement of the first problem is to calculate the distances on a map.

Many examples of geographic investigations have to use maps representation. Goggle maps is one of the platform which could help us to develop investigations of distances between two points on the map in order to analyze these on a bitmap image but the platform do not let us design points for the specified locations and do not give the possibility to have the result of a distance between fixed points. These problems have good solution on GeoGebra platform. In the map picture imported from Google we specify two locations: City of Braila and City of Galati in Romania, (fig.1). The image was fixed on the space of work of GeoGebra software and grid; axes were designed respecting the real scale of distances. The way on the route was established step by step using successive points and to create the possibility to have the sum of the distances indicating the real distance between the cities, (fig.2). If other maps will be investigated from this point of view, the probability to discover some anomalies is possible and this will indicate us that only one representation on the map will be real. Another solved problem is the possibility to change the visible map, changing the scale in a real time, quickly, without changing the representation of the imported map. A contextual control of the imported map picture using the specialized software of recovering the image is the opportunity to have many other analyses on the map when a big scale will be adopted, in order to have the opportunity to analyze small squares on the map. Unfortunately, more and more heterogeneous geographical data are created in the past years, which hindered the development of geographic modeling, model integration and analysis. So it is urgent to find a new solution, especially a new data representation strategy, for understanding, describing and representing geographic data perfectly. This will facilitate the integration of multi-sourced heterogeneous data which supports geographical problem solving in virtual geographical environments. It is useful and meaningful for geographic modeling and analysis, [Hon02].

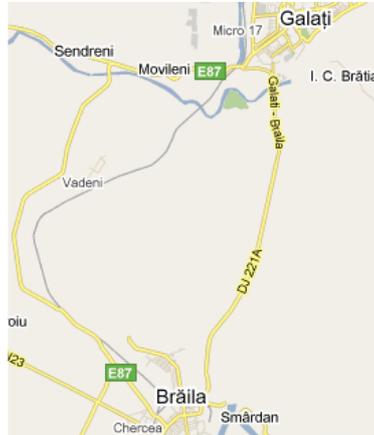

**Fig.1. Region of Brăila and Galaţi**

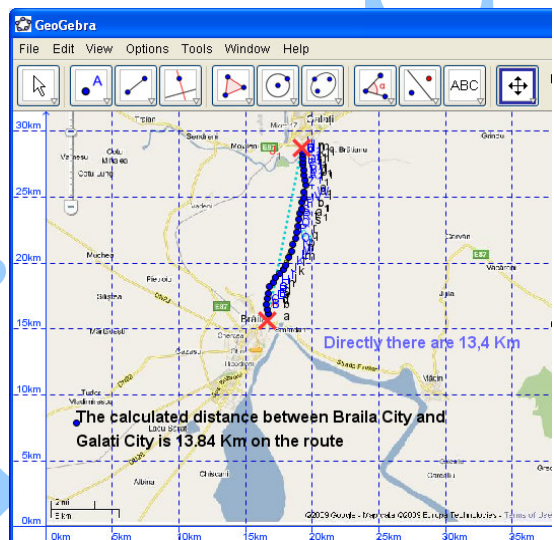

**Fig.2. GeoGebra in distance analysis**

The statement of the second problem is the dynamic analyze of the pollutant dispersion in surface waters.

In order to analyze the surface water qualities in Danube Delta we must create a dynamic map first according to the 2D topography of the region.

A map will be imported in the GeoGebra workspace and real distances data will be integrated in a good scale. The course of Sulina

Chanel and one of the interior lakes, Isacov, will be positioned with consecutive points in the work space, (fig. 4). The efficient understanding and application of mathematical models in the study of environmental phenomena keep up with the latest results in the mathematical domain, which could provide solutions for controlling, analyzing, predicting and studying of risk phenomena.

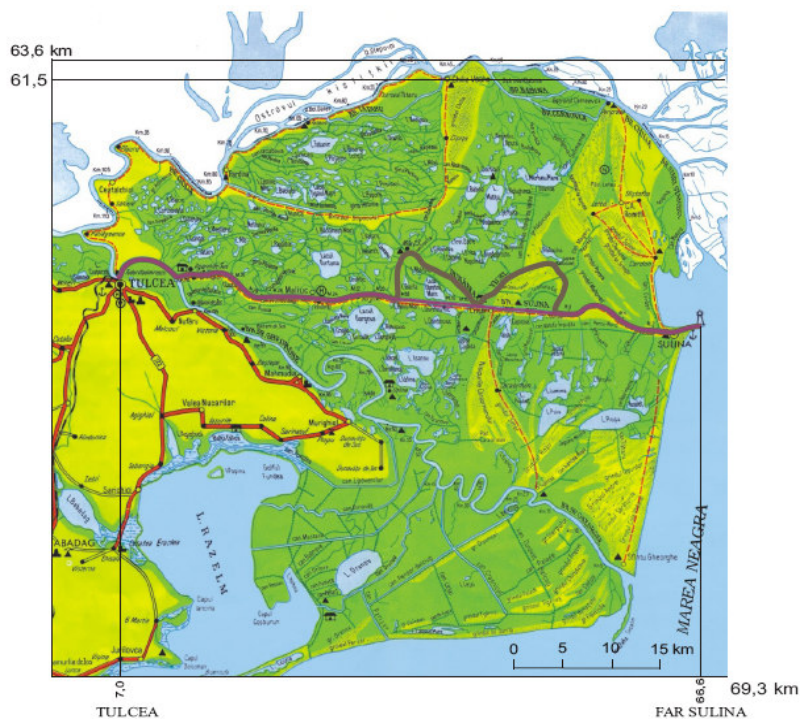

**Fig. 3. Danube Delta map**

Principles of mathematical shaping accept that we live in a world of models. Each life is a continuous confrontation between the shape of the ego and the model of the surrounding world. The majority of sciences resulted and developed significantly after some models had been previously elaborated. [Neu03]. The area of Isacov Lake is about 4.2x3.1=13.02 $Km^2$ and 16.33 $Km^2$ according to the Geogebra model, (fig.4). The area of Isacov Lake is about 16.33 $Km^2$ according to the Geogebra model which uses a step by step construction using points around, (fig.5).

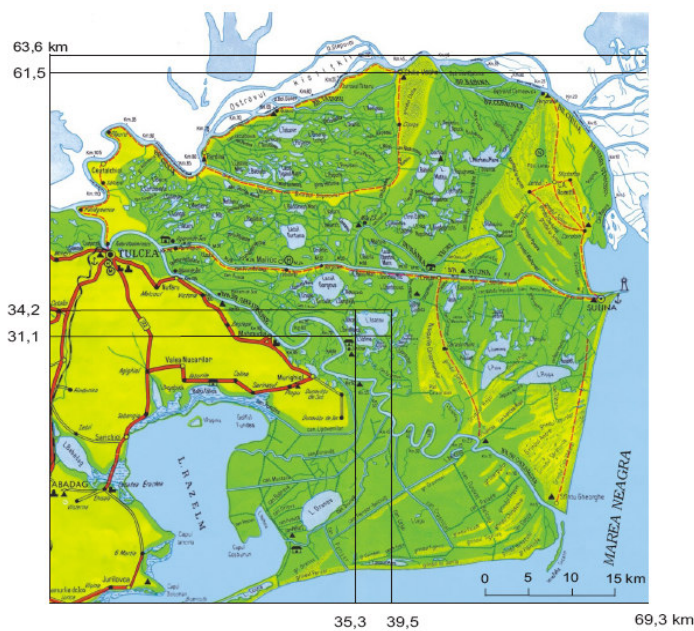

**Fig.4. Isacov Lake – map view**

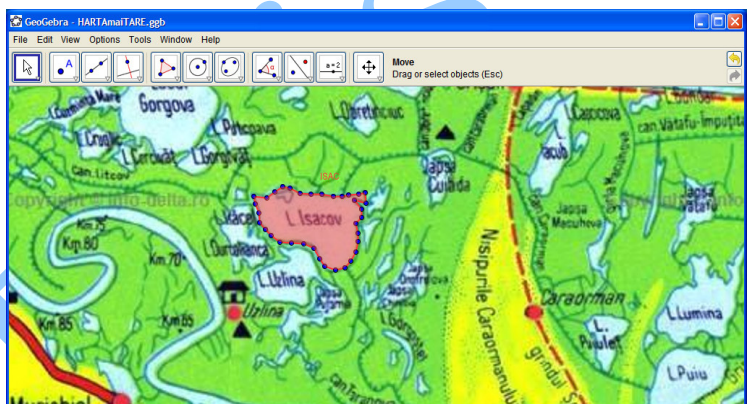

**Fig.5. Area of Isacov Lake – GeoGebra model**

In addition to these Geogebra projects, several other topics provide opportunities for software to be involved in other environmental research. The mix of geologists, engineers, and geographers has been essential to the development of new studies about groundwater flow and transport models, constructed to better exploit new findings in the environment engineering and to combine them with high-performance computing. For the study of water quality, many Groundwater Research Groups are interested in the development of high performance with computer models, to simulate the

177

movement of groundwater and contaminants at regional scales, as well as the application of such models to environmental management problems. For this a map of the situ is necessary, such as satellite maps that could be used fervently, (fig. 6).

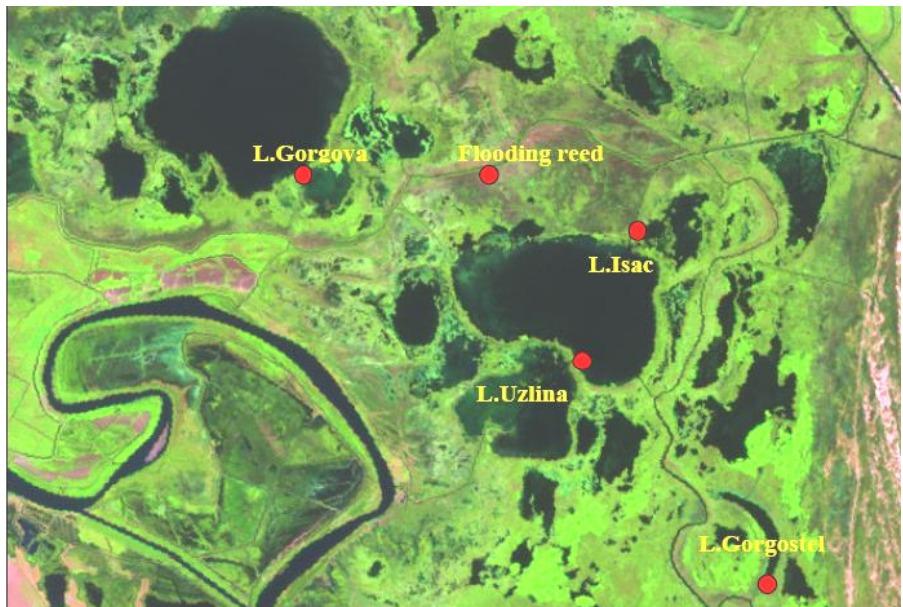

**Fig.6. Lakes in Danube Delta**

The paradigm of shaping a process acknowledges that an overview of the similar models, created for the issue under discussion, is a must. Sometimes the adjustment or the optimization of the already existing models is more than necessary. It is necessary that the useful information should be collected in order to settle and understand all the parameters of the problem. It is also recommendable firstly to design a simplified model to which details can be added which leads to achieving a flexible model, [Ing07].

Modeling projects of this nature are inherently multidisciplinary, involving expertise in hydrogeology and measurement (Geology), development and management of large environmental data sets (Geography), and the development and application of high-performance computational models (Environmental Engineering).

Analyzing the map imagine, one could see that the area represented has not a continuous frontier, (fig.7).

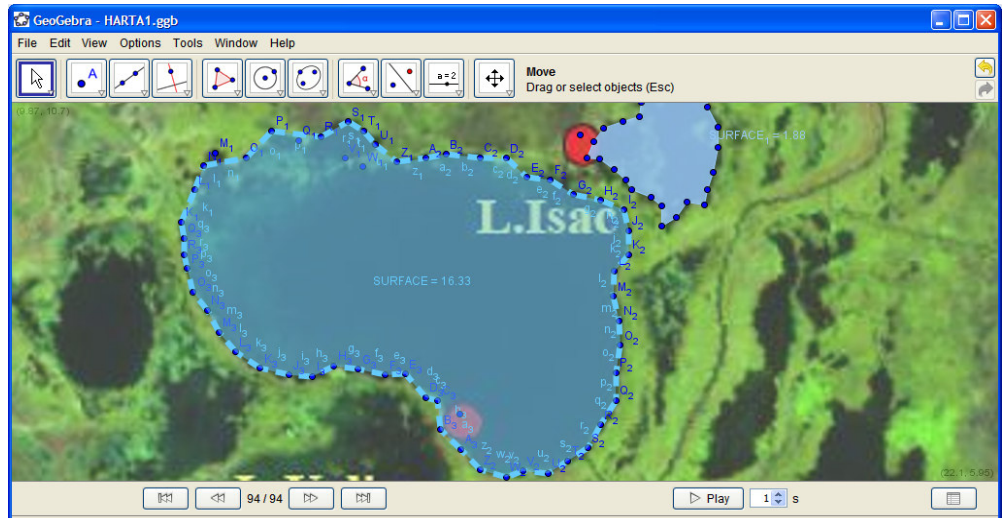

**Fig.7. Area of Isacov Lake step by step construction in GeoGebra**

This idea suggest that if the map is imported in the GeoGebra space of work and a continuous function which depends on "n" parameters is manipulated with some Geogebra commands, the dispersion and the representation will be better and the Geogebra will permit us to change the scale dynamically, without the damaging the map. This context allows us to have a virtual representation of the environment, closer to reality.

**Conclusions**

The special cooperation of Math, Geography and IT representations on the GeoGebra platform is one of the goals that were established in order to develop some future studies. A step by step construction, which represents the visual interpretation of the geographical context, could be a starting point for the future investigation. The problem of the transition from the 2D representation to the 3D representation with GeoGebra seems to be more interesting in this context, in order to solve problems focused on the analysis of the real topology of the environmental studies represented by 3D images.

One appreciates the pedagogical implications of exploring problems in geography in a dynamic environment, both as an investigation tool and as a demonstration tool, yet the collaboration between Math and Geography teachers and specialists in geography and in computers is a challenge. The term "Dynamic-Info-Geography" could be a new subject for GeoGebra-software developers. The mathematical model of the evolution of the Danube water quality parameters and the mathematical accepted structures can help to establish a more comprehensive map of the risk factors, the image of the complex system1 of Danube River (which lives like a human entity), being continuously monitored.